\documentstyle[aps,pre,epsf,preprint,eqsecnum]{revtex}
\tightenlines
\begin{document}
\title{Microstructural Shear Localization in Plastic Deformation of Amorphous Solids}
\author{J.S. Langer}
\draft
\date{\today}

\maketitle

\begin{abstract}
The shear-transformation-zone (STZ) theory of plastic deformation predicts that sufficiently soft, non-crystalline solids are linearly unstable against forming periodic arrays of microstructural shear bands. A limited nonlinear analysis indicates that this instability may be the mechanism responsible for strain softening in both constant-stress and constant-strain-rate experiments. The analysis presented here pertains only to one-dimensional banding patterns in two-dimensional systems, and only to very low temperatures.  It uses the rudimentary form of the STZ theory in which there is only a single kind of zone rather than a distribution of them with a range of transformation rates.  Nevertheless, the results are in qualitative agreement with essential features of the available experimental data. The nonlinear theory also implies that harder materials, which do not undergo a microstructural instability, may form isolated shear bands in weak regions or, perhaps, at points of concentrated stress. 
\end{abstract}
\bigskip

\section{Introduction}
\label{sec:intro}

In a recent publication \cite{FL}, M.L. Falk and I proposed a ``shear-transformation-zone" (STZ) theory of plastic deformation in amorphous solids.  This theory was based on direct observations of micro-scale rearrangements in molecular dynamics simulations.  The central feature of the theory was the recognition that the STZ's --- localized regions where irreversible molecular rearrangements occur --- are  two-state systems; they can switch back and forth between only two orientations.  Thus, the density of zones and the average of their orientations are appropriate order parameters for characterizing the internal states of such systems.  The equations of motion for these order parameters, including terms that describe annihilation and creation of STZ's during deformation, produce a relatively simple theory of plasticity that exhibits strain hardening, a dynamic transition between viscoelasticity and viscoplasticity at a yield stress, and hysteretic effects.

As presented in \cite{FL}, the STZ theory is strictly a mean field theory; it allows no spatial correlations in the stress or strain fields or in the new order parameter.  This assumption of spatial homogeneity, however, is inconsistent with the original picture when examined in closer detail.  A localized shear transformation produces a quadrupolar elastic displacement field in its neighborhood, and a corresponding change in the local stress pattern.  It is easy to see that this change in stress may induce shear transformations in neighboring zones along the directions of maximum shear.  (See Sections \ref{sec:Quasilinear} and \ref{sec:Stability} for details.)  If each localized shear transformation induces others along certain preferred directions, then the system as a whole may be unstable against formation of a pattern of shear bands.  This argument is the basis of numerical simulations described in an important series of papers by Bulatov and Argon.\cite{BA1,BA2,BA3}  The latter authors used a model of localized shear transformations that is more phenomenological than the STZ theory and that does not contain some of its central features, for example, the transition between viscoelasticity and viscoplasticity or the hysteretic effects.  One purpose of the present investigation has been to explore the Bulatov-Argon picture of microstructural shear banding in the context of the STZ theory.  

In its original mean-field version, the STZ theory predicts the time dependence of the plastic shear strain $\varepsilon^{pl}(t)$ in experiments (creep tests) where a uniform (deviatoric) stress $s$ is applied quickly and then held at a constant value.  For stresses less than an ultimate yield stress $s_y$, the strain approaches a value $\varepsilon^{pl}_{final}(s) =\varepsilon^{pl}(t\to \infty)$ and remains constant there; that is, the two-state systems become ``jammed."  The quantity $\varepsilon^{pl}_{final}(s)$ diverges as $s$ approaches $s_y$ from below, as does the characteristic time for approach to the final state. A graph of $s$ as a function of $\varepsilon^{pl}_{final}$ therefore looks like an ordinary stress-strain curve with plastic yielding, although the STZ interpretation of it as the infinite-time limit is not strictly conventional.  For applied stresses above $s_y$, the system comes to steady state at a nonzero strain rate $\dot\varepsilon^{pl}_{final}(s)$, which increases with increasing $s>s_y$.  In this case, the characteristic time for approach to steady state diverges as $s$ approaches $s_y$ from above.  

I have (so far) found only two examples of constant-stress experiments of this kind reported in the literature, an early one by Ender in 1968 \cite{ENDER} and another more recent investigation by Hasan and Boyce in 1995\cite{HB}, both using the polymer PMMA.  The work of Hasan and Boyce (HB) is rich in detail and is supplemented by a theoretical analysis of their data.  At small applied shear stresses (up to about $80\%$ of what I interpret to be an ultimate yield stress), the HB results seem to be consistent with the mean-field STZ prediction, i.e. the strain rate decreases monotonically to zero.  At higher stresses, however, the curves of strain versus time have inflection points where the strain rate goes through a minimum and then increases, indicating the onset of some new mode of viscoplastic deformation. Significantly, the transition from viscoelastic to viscoplastic response occurs within a very narrow range of applied stresses; that is, the strain rate is highly sensitive to the stress within this range.  I propose that this new mode of deformation may emerge from a microstructural shear-banding instability.  

The most common measurements of plastic deformation are performed under conditions of controlled strain, specifically, at constant strain rates. For soft materials such as polymeric glasses or soils, stress-strain curves generated in this way often exhibit strain softening. That is, the stress initially rises with increasing strain but then goes through a maximum and decreases, so that the stress ultimately required to sustain the fixed strain rate remains lower than the peak stress. Hasan and Boyce\cite{HB} show clear examples of this kind of behavior; similar data for soils can be found in the book by Ishihara\cite{ISHIHARA}.  The converse of the strong stress sensitivity seen in the constant-stress experiments is the fact that the peak stresses in the constant-strain-rate experiments are only weakly dependent on the strain rate. I suggest that this version of strain softening also can be understood as the result of an instability that sets in near the peak stress and produces a pattern of microstructural shear bands which account for the deformation rate in the subsequent steady state.  

The theoretical analysis of Hasan and Boyce\cite{HB} provides important background for the present investigation.  Like the STZ theory, the HB theory is based on a picture of localized shear-transformation sites whose populations and states of deformation are the internal state variables that characterize the system. Unlike STZ, however, HB do not try to model the molecular rearrangements taking place at their sites or the interactions between these sites; thus they do not attempt to discover a connection between strain softening and shear localization.  Nevertheless, there are deep relations between these theories.  Both are attempts to identify the physically relevant internal variables whose equations of motion must ultimately be the basis of any satisfactory, history-dependent theory of plasticity --- as opposed to, say, purely phenomenological theories based on yield surfaces\cite{LUBLINER} or Masing rules\cite{MASING}.  (Unfortunately, HB use the strain itself as a state variable, which cannot be correct from a fundamental point of view.)  Moreover, both theories identify cumulative internal changes during plastic deformation as the basis for dynamic yielding at high stress.

The scheme of this paper is as follows. Section \ref{sec:STZ} contains a brief outline of the basic ideas of the STZ theory and serves to define concepts and  notation for the rest of the analysis. Some parts of the discussion presented there have not been published before now. In Section \ref{sec:Quasilinear}, I derive an approximate, quasilinear, but fully tensorial generalization of the STZ equations that allows me to consider spatially inhomogeneous stress changes.  Section \ref{sec:Stability} contains a linear stability analysis of the spatially uniform steady-state solutions of the quasilinear theory. The expected instabilities emerge, but with some surprises. The quasilinear analysis provides the basis for a nonlinear integro-differential equation that describes, in an effectively one-dimensional situation, how shear patterns may emerge in response to shear stresses.  The solutions of the latter equation exhibit the strain-softening behaviors described in the preceding paragraphs. They also provide a mechanism for the formation of isolated shear bands in materials that do not exhibit strain softening. The nonlinear analysis, including numerical results, is described in Section \ref{sec:Nonlinear}.

The results presented here are, at best, a rough caricature of plastic deformation in shear-banding or strain-softening materials.  Some features seem to be quite convincing, but others remain problematic.  Answers to several central questions, especially the importance of the one-dimensional approximation made in Section \ref{sec:Nonlinear}, will require numerics and laboratory experiments that are beyond the scope of the present investigation.  

\section{Elements of the STZ Theory}
\label{sec:STZ}

As in \cite{FL}, the discussion presented here pertains only to two-dimensional non-crystalline systems at effectively zero temperatures, i.e. at temperatures far below the glass transition. Also, following \cite{FL}, it is useful to start with the special case of a single preferred orientation for the zones and the applied stress, and to denote the number-density of zones oriented in the ``$+$''/``$-$'' directions by the symbol $n_{\pm}$.  Unlike \cite{FL}, however, the convention here is one in which the externally applied deviatoric stress is diagonal in the $x$, $y$ coordinates, specifically $s_{xx}=-s_{yy}=s_0$, and $s_{xy}=0$. Choose the ``$+$" zones to be oriented (elongated) along the $y$ axis, and the ``$-$" zones along the $x$ axis. With this convention, we can --- temporarily --- suppress tensor notation.  

As in \cite{FL} Eq.(3.10), the plastic strain rate is:
\begin{equation}
\dot\varepsilon^{pl}=b^2\,(\Delta\varepsilon)\,W
\end{equation}
where $b^2$ (previously denoted $V_z$) is the typical area of an STZ, $(\Delta\varepsilon)$ is the increment of local shear strain that occurs when one of these zones undergoes a transition (a number of order unity), and
\begin{equation}
W=R_+n_+-R_-n_-.
\end{equation}
Here, $R_+$ and $R_-$ are the rates for $``+"\to ``-"$ and $``-"\to ``+"$ transitions respectively. 
These transition rates, at very low temperatures, are entropy-activated as opposed to thermally activated:
\begin{equation}
\label{ratefactor}
R_{\pm}(s_0)= R_0\,\exp\left[-{\Delta V^*(\pm s_0)\over v_f}\right],
\end{equation}
where $\Delta V^*(\pm s_0)$ is the excess free volume (area) needed to allow a $``\pm" \to ``\mp"$ transition to occur, and $v_f$ is the average free volume in the system.  In \cite{FL}, $\Delta V^*(\pm s_0)$ was chosen to have the form
\begin{equation}
\Delta V^*(s_0) =V_0^*\,\exp(-s_0/\bar\mu),
\end{equation}
where $V_0^*$ is of order the average volume per molecule and $\bar\mu$ is generally much smaller than the shear modulus.  $R_0$ is an attempt frequency that, in \cite{FL}, was found to be proportional to the square root of the total strain rate.  Here, I shall assume that this factor is simply a constant. 

If there is no spatial dependence of any of the variables, then the equation of motion for $n_{\pm}$ is
\begin{equation}
\dot n_{\pm}=R_{\mp}n_{\mp}-R_{\pm}n_{\pm}+(\dot\varepsilon^{pl}\,s_0)\Bigl[A_{cr}-A_{ann}\,n_{\pm}\Bigr].
\end{equation}
This equation includes terms which describe creation and annihilation of STZ's at rates proportional to the rate at which plastic work is being done on the system, $\dot\varepsilon^{pl}\, s_0$, with coefficients $A_{cr}$ and $A_{ann}$ respectively.

As previously, define $n_{\Delta}=n_--n_+$, $n_{tot}=n_-+n_+$. Also,
\begin{equation}
R_{\pm}(s_0)=R_0\,\left[{\cal C}(s_0)\pm{\cal S}(s_0)\right],
\end{equation}
where
\begin{eqnarray}
{\cal C}(s_0)&=&\exp\left[-{V_0^*\over v_f}\,\cosh\left({s_0\over\bar\mu}\right)\right]\, \cosh\left[{V_0^*\over v_f}\,\sinh\left({s_0\over\bar\mu}\right)\right],\nonumber\cr\\ {\cal S}(s_0)&=&\exp\left[-{V_0^*\over v_f}\,\cosh\left({s_0\over\bar\mu}\right)\right]\, \sinh\left[{V_0^*\over v_f}\,\sinh\left({s_0\over\bar\mu}\right)\right].
\end{eqnarray}
Then
\begin{equation}
\dot n_{\Delta} = 2R_0\,{\cal S}(s_0)\,n_{tot}-2R_0\,{\cal C}(s_0)\,n_{\Delta} - (\dot\varepsilon^{pl}\, s_0)\,A_{ann}\,n_{\Delta};
\end{equation}
and 
\begin{equation}
\label{dotntot}
\dot n_{tot}=(\dot\varepsilon^{pl}\, s_0)\left[2\,A_{cr}-A_{ann}\,n_{tot}\right]
\end{equation}

Eq.(\ref{dotntot}) implies that the ratio $2A_{cr}/A_{ann}\equiv n_{\infty}$ is the natural equilibrium value for $n_{tot}$, the total density of STZ's.  In principle, $n_{\infty}$ could be a stress dependent quantity.  Amorphous materials (even the Lennard-Jones solid in the molecular dynamics simulations described in \cite{FL}) generally exhibit a second order dilation in response to a shear stress; thus the equilibrium density of STZ's might increase as a function of $s_0^2$.  For simplicity, however, I shall not consider that possibility here.

Next, transform to dimensionless variables: $n_{\Delta}=n_{\infty}\,\Delta$, $n_{tot}=n_{\infty}\,\Lambda$, and write
\begin{equation}
\label{epsdot}
\dot\varepsilon^{pl}={\varepsilon_0\over \tau(s_0)}\,w,
\end{equation}
where
\begin{equation}
w = {\cal T}(s_0)\,\Lambda-\Delta;
\end{equation}
\begin{equation}
\label{parameterdefs}
\varepsilon_0 = {1\over 2}\,b^2\,(\Delta\varepsilon)\,n_{\infty};~~~~{1\over\tau(s_0)}= 2\,R_0\,{\cal C}(s_0);
\end{equation}
and
\begin{equation}
\label{Tdef}
{\cal T}(s_0)=\tanh\,\left[{V_0^*\over v_f}\,\sinh\left({s_0\over\bar\mu}\right)\right].
\end{equation} 
The equations of motion for $\Delta$ and $\Lambda$ become
\begin{equation}
\label{deltadot}
\dot\Delta = {w\over \tau(s_0)}\,\left(1 - \gamma\,s_0\,\Delta\right) 
\end{equation}
and
\begin{equation}
\label{lambdadot}
\dot \Lambda = {\gamma\over \tau(s_0)}\,w\,s_0\,(1-\Lambda).
\end{equation}
Here, $\gamma=(1/2)\,A_{ann}\,b^2\,(\Delta\varepsilon)\,n_{\infty}$, which I shall assume to be a stress-independent quantity.

As pointed out in \cite{FL} (see also \cite{LL1}, \cite{LL2} and \cite{FLMRS}), Eqs.(\ref{deltadot}) and (\ref{lambdadot}) have two kinds of  solutions, one --- the ``jammed'' or ``viscoelastic'' solution --- in which the plastic rate of deformation $\dot\varepsilon^{pl}$ vanishes in steady state, and another --- the ``viscoplastic'' solution --- in which it does not. In the first case, there is a one-parameter family of solutions with $\Delta = \Delta_0= {\cal T}(s_0)\,\Lambda_0$ where $\Lambda=\Lambda_0$ can have a continuous range of values determined dynamically by the density of STZ's in the initial state of the system.  In the second, viscoplastic case, we can have only $\Delta = \Delta_0= 1/(\gamma\,s_0)$ and $\Lambda = \Lambda_0 = 1$.  The jammed states are dynamically stable attractors for values of $s_0$ less than an ultimate yield stress $s_y$, determined by
\begin{equation}
\label{yieldstress}
\gamma\,s_y\,{\cal T}(s_y)=1.
\end{equation}
Conversely, the viscoplastic state is stable for $s_0>s_y$. To see these stability properties, write $\Delta = \Delta_0+\Delta_1(t)$, $\Lambda = \Lambda_0+\Lambda_1(t)$, and compute the amplification rate $\omega_0(s_0)$ for which $\dot\Delta_1=\omega_0\,\Delta_1$, $\dot\Lambda_1=\omega_0\,\Lambda_1$. The result is that, for $s_0<s_y$, the two eigenvalues are  $\omega_0=0$ (because $\Lambda_0$ is undetermined), and $\omega_0=\Omega_0$, where
\begin{equation}
\Omega_0(s_0)=2\,{\cal C}(s_0)\,\Bigl[\gamma\,s_0\,{\cal T}(s_0)-1\Bigr].
\end{equation}
For $s_0>s_y$, the $\Delta_1$ and $\Lambda_1$ modes are decoupled, and both have the eigenvalue $\omega_0=-\Omega_0$.  Because $\Omega_0$ is an increasing function of $s_0$ that vanishes at $s_y$, the stable attractor crosses from the viscoelastic to the viscoplastic branch of steady states at $s_0=s_y$.

\section{Quasilinear Theory of Inhomogeneous Stress Patterns}
\label{sec:Quasilinear}

So far, this analysis has pertained only to systems in which the shear stress has magnitude $s_0$ everywhere and has only a fixed orientation. To incorporate inhomogeneous stress tensors $s_{ij}$ with varying orientations into this analysis, we need --- at least temporarily --- a fully tensorial version of the theory.  No such version of the complete STZ theory exists at present.  (See Appendix A of Falk's dissertation \cite{Fthesis} for a first attempt at developing such a version.)  Therefore, I shall resort to a quasilinear approximation \cite{LL1,FLMRS} in order to derive equations of motion for the inhomogeneous stress variations. 

Start by noting that the (deviatoric) stress and plastic strain must in general be traceless, symmetric tensors, $s_{ij}$ and $\varepsilon^{pl}_{ij}$ respectively.  It follows from the form of Eqs.(\ref{epsdot}), (\ref{deltadot}), and (\ref{lambdadot}) that $\Delta$, $w$, and ${\cal T}$  also must generalize to traceless, symmetric tensors, and that $\Lambda$ must be a scalar.  One way to obtain such a structure is to write
\begin{equation}
\label{Tapprox}
{\cal T}_{ij}\approx \lambda\,s_{ij},~~~~\lambda \approx V_0^*/(v_{f0}\bar\mu);
\end{equation}
that is, linearize the right-hand side of Eq.(\ref{Tdef}) in $s_0$ and then let $s_0$ become $s_{ij}$.  To be consistent, we also must approximate the time constant $\tau(s_0)$ by its limit at small $s_0$, which I denote simply by $\tau_0$:
\begin{equation}
\label{tauapprox}
{1\over\tau_0}\approx 2\,R_0\,e^{-V_0^*/v_{f0}}.
\end{equation}
These estimates for $\lambda$ and $\tau_0$, however, should not be taken literally; they represent only one of possibly many ways in which we might start from atomistic considerations to derive a theory with the general phenomenological structure shown below.  

With the approximations (\ref{Tapprox}) and (\ref{tauapprox}), we now have:
\begin{equation}
\label{epsdottensor}
\dot\varepsilon^{pl}_{ij}= {\varepsilon_0\over\tau_0}\,w_{ij},~~~~w_{ij}=\lambda\,s_{ij}\,\Lambda-\Delta_{ij}.
\end{equation}
Also, (using summation convention),
\begin{equation}
\label{deldottensor}
\dot\Delta_{ij}={1\over\tau_0}\,\left[w_{ij}- {1\over 2}\,\gamma\,(w_{kl}\,s_{kl})\,\Delta_{ij}\right]
\end{equation}
and
\begin{equation}
\label{lambdadottensor)}
\dot \Lambda = {1\over 2\,\tau_0}\,(w_{kl}\,s_{kl})\,\gamma\,(1-\Lambda). 
\end{equation}
The dissipation rate, denoted by $\dot\varepsilon^{pl}\,s_0$ in Section \ref{sec:STZ}, generalizes here to $(1/2)\,\dot\varepsilon^{pl}_{kl} \,s_{kl}$.  Note that, if we  guess that $\Lambda\to 1$ quickly and stably and therefore set $\Lambda = 1$ in Eq.(\ref{epsdottensor}), we obtain the quasilinear STZ equations that were used in \cite{LL1} and \cite{LL2}.  If we did this here, however, we would miss some of the most important features of the shear-banding mechanism.

The new element of the present theory is that the stress tensor $s_{ij}$ is to become a dynamical quantity varying with both position and time ($x$, $y$, and $t$).  Before writing  equations of motion for $s_{ij}$, I must specify the physical picture that I have in mind.  The molecular dynamics simulations reported in \cite{FL} imply that the STZ's are sparsely distributed sites at which small clusters of atoms undergo irreversible rearrangements in response to applied stresses.  To visualize an STZ, think of a void in an elastic material, and place a small group of atoms inside this void in such a way that their average free volume is somewhat larger than that for most other atoms in the system.  The void has some degree of structural stability; it can deform elastically but, because of the configuration of atoms on its surface, it resists collapse.  Rearrangements of the atoms that are caged within the void couple to its shape and, therefore, to the stress field in the elastic medium in which the void is embedded.  (The stress field outside the void is similar to that in the neighborhood of an Eshelby inclusion. \cite{ESHELBY}) This stress field --- outside the STZ's --- provides the mechanism by which the STZ's are coupled to each other. 

To distinguish it from the uniform applied stress $s_0$ (as defined at the beginning of Section \ref{sec:STZ}), denote this fluctuating part of the stress field by the symbol $\tilde s_{ij}(x,y,t)$.  For simplicity, first consider a $``-"\to ``+"$ STZ transition located at the origin of the coordinate system and aligned along the original $x,y$ axes; and compute the stresses and strains in its not-too-close neighborhood by the potential method.  That is, define $\Phi(x,y)$ such that
\begin{equation}
\label{potential1}
\nabla^2\Phi=0;~~~~u_x=\partial_x\Phi;~~~~u_y=\partial_y\Phi;
\end{equation}
and choose
\begin{equation}
\label{potential2}
\Phi(x,y)\sim {y^2-x^2\over(x^2+y^2)^2}={\cos 2\theta\over r^2}
\end{equation}
so that $u_x\sim x^{-3}$ along the $x$ axis and $u_y\sim -y^{-3}$ along the $y$ axis.  Then the deviatoric stress and strain fields have the form
\begin{equation}
\label{stresspattern}
\tilde s_{xx}-\tilde s_{yy}\sim\partial_x^2 \Phi-\partial_y^2 \Phi \sim -{x^4-6x^2y^2+y^4\over (x^2+y^2)^4}=-{\cos 4\theta\over r^4},
\end{equation}
where $r$ and $\theta$ are the conventional polar coordinates. This function has negative lobes along the $x$ and $y$ axes and positive lobes along the lines $x=\pm y$, so that the shear in the direction of the initial deformation is largest along the diagonals.  

Now suppose that the STZ transformation which caused the stress pattern (\ref{stresspattern}) is rotated by an angle $\vartheta$ relative to the $x$, $y$ coordinate system.  Then
\begin{equation}
\Phi(x,y)={1\over r^4}\,[(y^2-x^2)\cos 2\vartheta - 2xy\,\sin 2\vartheta] =-{r_i\,r_j\over r^4}\,{\cal D}_{ij}(\vartheta),
\end{equation}
where ${\cal D}$ is proportional to the director matrix: 
\begin{equation}
{\cal D}(\vartheta)={\rm constant}\times\pmatrix{\cos 2\vartheta & \sin 2\vartheta\cr \sin 2\vartheta & -\cos 2\vartheta}.
\end{equation}

If, according to (\ref{potential1}), $\Phi$ is the potential for displacements, then, for any position ${\bf r}$ far enough away from the origin that it is outside the transforming zone,
\begin{equation}
\dot {\tilde s_{ij}}({\bf r}) = 2\mu\,\dot\varepsilon^{el}_{ij}({\bf r})=-2\mu\,b^4\,\partial_i\,\partial_j\,{r_k\,r_l\over r^4} \,\dot\varepsilon_{kl}^b(0).
\end{equation} 
Here, $\dot\varepsilon^{el}_{ij}({\bf r})$ is the elastic strain rate induced at ${\bf r}$, $\mu$ is the elastic shear modulus, and $b$ is the characteristic size of an STZ.  The quantity $\dot\varepsilon_{kl}^b(0)$ is the strain rate induced by the shear transformation at distances $r$ of order $b$, which, in turn, is proportional to the rate $\dot n_{\Delta}(0)$
at which the STZ population is changing at the point ${\bf r}=0$.  More generally, the source of the fluctuating stress field $\dot {\tilde s_{ij}}({\bf r})$ has a density distribution at a point, say, ${\bf r}'$ given by
\begin{equation}
n_{tot}\,\dot\varepsilon_{kl}^b({\bf r}')= (\Delta\varepsilon)\,\dot n_{\Delta}({\bf r}')=
(\Delta\varepsilon)\,n_{\infty}\,\dot \Delta_{kl}({\bf r}').
\end{equation}
Integrating over the population of STZ's with this source density, we find an equation of motion for the deviatoric stress field:
\begin{equation}
\label{sdoteq}
\dot{\tilde s}_{ij}({\bf r})=-2\mu\,b^4\,(\Delta\varepsilon)\,n_{\infty}\,\int d^2r'\,\left[{\partial\over\partial r_i}\,{\partial\over\partial r_j}\,{(r_k-r_k')\,(r_l-r_l') \over |{\bf r}-{\bf r}'|^4}\right]\,\dot\Delta_{kl}({\bf r}')\,H(|{\bf r}-{\bf r}'|-b).
\end{equation}
Here, $H$ is the Heaviside function that cuts off the integration inside the core of the STZ.  (It may be more realistic, but probably not essential, to use a smoother cutoff.) Because this stress field comes directly from Eqs.(\ref{potential1}) and (\ref{potential2}), it automatically satisfies the force balance and compatibility conditions.

Eq.(\ref{sdoteq}) is best expressed in terms of the Fourier transform of $\tilde s_{ij}(x,y)$, say $\hat s_{ij}({\bf k})$. Then,
\begin{eqnarray}
\label{shat12}
\nonumber
{1\over 2\mu\,(\Delta\varepsilon)\,b^2\,n_{\infty}}\,\dot{\hat s}_{xx}({\bf k}) &=& \hat{\cal G}_1({\bf k})\, \hat {\dot\Delta}_{xx}({\bf k})+ \hat{\cal G}_2({\bf k})\,\hat {\dot\Delta}_{xy}({\bf k});\cr\\
{1\over 2\mu\,(\Delta\varepsilon)\,b^2\,n_{\infty}}\,\dot{\hat s}_{xy}({\bf k}) &=& \hat{\cal G}_2({\bf k})\, \hat{\dot\Delta}_{xx}({\bf k})- \hat{\cal G}_1({\bf k})\,\hat{\dot\Delta}_{xy}({\bf k}).
\end{eqnarray}
Note that these are linear relations between the time derivatives of the fluctuating part of the stress field and the order parameter $\Delta$.  

The kernels $\hat{\cal G}$ contain much of the dynamical information that we need.  Write the wavevector ${\bf k}$ in the form $k_x=k\cos\psi$, $k_y=k\sin\psi$.  Then
\begin{equation}
\hat{\cal G}_1(k,\psi)=-\cos (4\psi)\,\hat g(bk);~~~~ \hat{\cal G}_2(k,\psi)= -\sin (4\psi)\,\hat g(bk);
\end{equation}
and
\begin{eqnarray}
\label{ghat}
\hat g(\kappa)&=&6\,\kappa^2\,\int_{\kappa}^{\infty}{d\rho\over \rho^3}\,\int_0^{2\pi}d\theta\, e^{-i\rho\cos\theta}\,\cos 4\theta \cr\nonumber\\ &=& 12\,\pi\,\kappa^2\,\int_{\kappa}^{\infty}{d\rho\over \rho^3}\,J_4(\rho)= {12\pi\over\kappa}\,J_3(\kappa),
\end{eqnarray}
where $\kappa=bk$ and the $J$'s are Bessel functions. Therefore, $\hat g(\kappa)$ approaches $\pi\kappa^2/4$ as $\kappa \to 0$ and is proportional to $\kappa^{-3/2}$ as $\kappa\to\infty$. It rises through a maximum of about 4.176 at $\kappa=\kappa_c\cong 3.611$ and then oscillates as it decreases. In the following, I denote the maximum value of $\hat g(\kappa)$ by $\nu_c^{-1}$, i.e. $\nu_c \cong 0.2395$.

\section{Stability of the Uniform Steady States in the Quasilinear Theory}
\label{sec:Stability}

The next step is use the results of the preceding Section to examine the linear stability of spatially uniform steady-state solutions of our equations of motion. As we shall see in Section \ref{sec:Nonlinear}, the linear stability analysis of steady states misses crucially important transient phenomena, but it does give us some insight about the behavior of this system.  

It is convenient to use a notation in which
\begin{equation}
\label{sij}
s_{ij}=\pmatrix{s_0&0\cr 0&-s_0}+\pmatrix{\hat s_1&\hat s_2\cr \hat s_2&-\hat s_1}\,e^{i{\bf k}\cdot {\bf r} +\omega t},
\end{equation}
with analogous notation for components of $\Delta$, and
\begin{equation}
\Lambda =\Lambda_0 + \hat\Lambda_1\,e^{i{\bf k}\cdot {\bf r} +\omega t}.
\end{equation}
The second term on the right-hand side of Eq.(\ref{sij}) is the fluctuating stress tensor $\tilde s_{ij}$.  Like all other first order terms in this linear analysis, its time dependence is $\exp(\omega t)$. 

At zero'th order, according to Eqs. (\ref{epsdottensor}), (\ref{deldottensor}), and (\ref{lambdadottensor)}), we have
\begin{equation}
\label{del0eqn}
\dot\Delta_0={1\over\tau_0}\,(\lambda s_0\,\Lambda_0-\Delta_0)\,(1-\gamma\,s_0\,\Delta_0),
\end{equation}
and
\begin{equation}
\label{lam0eqn}
\dot\Lambda_0={\gamma\,s_0\over\tau_0}\,(\lambda s_0\,\Lambda_0-\Delta_0)\, (1-\Lambda_0).
\end{equation}
In this quasilinear approximation, the ultimate yield stress $s_y$, determined by Eq.(\ref{yieldstress}), is $s_y=1/\sqrt{\gamma\,\lambda}$. 
 
Consider first the jammed situation for which $s_0<s_y$, and examine the linear stability of the stationary state with $\Delta_0=\lambda\,s_0\,\Lambda_0$.  $\Lambda_0$ remains a free parameter to be determined by initial conditions.  (We shall compute $\Lambda_0$ explicitly in Section \ref{sec:Nonlinear}.) 

The first-order equations are the following:
\begin{equation}
\omega\,\tau_0\,\hat \Delta_1=(\lambda\,\Lambda_0\,\hat s_1+\lambda\,s_0\,\hat\Lambda_1 -  \hat\Delta_1)\,(1-\gamma\,s_0\,\Delta_0);
\end{equation}
\begin{equation}
\omega\,\tau_0\,\hat\Delta_2=\lambda\,\Lambda_0\,\hat s_2-\hat\Delta_2;
\end{equation}
and
\begin{equation}
\omega\,\tau_0\,\hat\Lambda_1= \gamma\,s_0\,(\lambda\,\Lambda_0\,\hat s_1+\lambda\,s_0\,\hat\Lambda_1 -  \hat\Delta_1)\,(1-\Lambda_0).
\end{equation}
From these relations, we find that
\begin{equation}
\label{delta12}
\hat\Delta_1={\lambda\,\Lambda_0\,D_{\Lambda}(s_0)\over \omega\,\tau_0 + D(s_0)}\,\hat s_1;~~~~
\hat\Delta_2={\lambda\,\Lambda_0\over \omega\,\tau_0 +1}\,\hat s_2;
\end{equation}
where
\begin{equation}
D(s_0)=1 - \left({s_0\over s_y}\right)^2;~~~~D_{\Lambda}(s_0)= 1 - \Lambda_0\,\left({s_0\over s_y}\right)^2.
\end{equation}

The combination of Eqs.(\ref{shat12}) and (\ref{delta12}) is a pair of linear, homogeneous equations that determines the amplification rate $\omega(k,\psi)$:
\begin{equation}
\label{eigen1}
\hat s_1 = -\cos (4\psi)\,\nu_{\infty}\,\hat g(bk)\,\left({D_{\Lambda}(s_0)\,\Lambda_0\over \omega\,\tau_0+D(s_0)}\right)\,\hat s_1 -\sin (4\psi)\,\nu_{\infty}\,\hat g(bk)\,\left({\Lambda_0\over \omega\,\tau_0+1}\right)\,\hat s_2;
\end{equation}
and
\begin{equation}
\label{eigen2}
\hat s_2 = -\sin (4\psi)\,\nu_{\infty}\,\hat g(bk)\,\left({D_{\Lambda}(s_0)\,\Lambda_0\over \omega\,\tau_0+D(s_0)}\right)\,\hat s_1 + \cos (4\psi)\,\nu_{\infty}\,\hat g(bk)\,\left({\Lambda_0\over \omega\,\tau_0+1}\right)\,\hat s_2.
\end{equation}
Here, 
\begin{equation}
\nu_{\infty}=2\mu\,\lambda\,(\Delta\varepsilon)\,b^2\,n_{\infty}.
\end{equation}
As we shall see, $\nu_{\infty}$ is the dimensionless group of parameters that controls stability against microstructural shear banding.

To understand the nature of the stability spectrum, it is easiest to look at two special cases.  The most important situation for our purposes is the orientation $\psi=\pm\pi/4$, where the bands are aligned with the direction of maximum applied shear stress.  In this case, the $\hat s_1$ and $\hat s_2$ modes decouple, and we have:
\begin{equation}
\label{omega45}
\omega(k,\pi/4)\,\tau_0 =\cases{\nu_{\infty}\,\hat g(bk)\,\Lambda_0\,D_{\Lambda}(s_0)-D(s_0)&($\hat s_1$ mode)\cr -\nu_{\infty}\,\hat g(bk)\,\Lambda_0-1&($\hat s_2$ mode).}
\end{equation}
As might be expected, an instability (positive $\omega$) can occur in the $\hat s_1$ mode where the shear deformation is parallel to the bands; the $\hat s_2$ mode, where the shear is normal to the bands, always decays.  Suppose, for the moment, that $\Lambda_0\to 1$ during the initial transient --- a possible consequence of the dynamics described by Eq.(\ref{lam0eqn}). Then, for the $\hat s_1$ mode in Eq.(\ref{omega45}), 
\begin{equation}
\label{omegaslow}
\omega\,\tau_0\to\left(1-{s_0^2\over s_y^2}\right)\,\Big[\nu_{\infty}\,\hat g(bk)-1\Big].
\end{equation}
There is an instability for all values of $s_0<s_y$, but only for $\nu_{\infty} > \nu_c\cong 0.2395$. When the latter condition is satisfied, the instability occurs in some range of values of $bk$ in the neighborhood of $\kappa_c \cong 3.611$. (The numerical constants $\nu_c$ and $\kappa_c$ are defined following Eq.(\ref{ghat}).) Note that the first factor on the right-hand side of Eq.(\ref{omegaslow}) vanishes as $s_0\to s_y$, implying that the instability weakens at large stresses and disappears at the yield stress.  

Unexpectedly (for me), the instability is stronger in the orientation $\psi=0$, where the $\hat s_1$ and $\hat s_2$ modes again decouple, but the bands are at forty-five degrees from the direction of maximum applied shear stress, i.e. in the direction in which the applied shear stress vanishes. We have
\begin{equation}
\label{omega0}
\omega(k,0)\,\tau_0 =\cases{-\nu_{\infty}\,\hat g(bk)\,\Lambda_0\,D_{\Lambda}(s_0)-D(s_0)&($\hat s_1$ mode)\cr \nu_{\infty}\,\hat g(bk)\,\Lambda_0-1&($\hat s_2$ mode).}
\end{equation}
In this case, only the $\hat s_2$ mode can be unstable.  The shear deformation is again parallel to the bands, but there is no weakening factor analogous to that in Eq.(\ref{omegaslow}) for this mode, and therefore it will grow more rapidly.  Apparently, at $\psi=\pi/4$, a substantial fraction of the STZ's are already aligned with the applied stress by the time that the system has reached its uniform steady state, and thus fewer zones are available to be transformed by the induced stress $\hat s_1$.  No such suppression occurs at $\psi = 0$, because there the bias in the orientation of the zones is not aligned with the orientation of the emerging bands.  

To complete this analysis, look at the viscoplastic states with $s_0>s_y$.  In steady state with nonzero plastic strain rate, $\Delta_0=1/(\gamma\,s_0)$ and, without ambiguity, $\Lambda_0=1$. In place of Eqs.(\ref{delta12}), we have:
\begin{equation}
\hat\Delta_1=-{\lambda\,(s_y^2/s_0^2)\,\bar D(s_0)\over \omega\,\tau_0 + \bar D(s_0)}\,\hat s_1,~~~~\hat\Delta_2={\lambda\over \omega\,\tau_0 + (s_0/s_y)^2}\,\hat s_2,
\end{equation}
where
\begin{equation}
\bar D(s_0)= {s_0^2\over s_y^2}-1.
\end{equation}
The analogs of Eqs.(\ref{eigen1}) and (\ref{eigen2}) are:
\begin{eqnarray}
\label{eigen1vp}
\nonumber
\hat s_1 = \cos (4\psi)\,\nu_{\infty}\,\hat g(bk)\,&&\left({\bar D(s_0)\,(s_y/s_0)^2\over \omega\,\tau_0+\bar D(s_0)}\right)\,\hat s_1\cr \\&&  -\sin (4\psi)\,\nu_{\infty}\,\hat g(bk)\,\left({1\over \omega\,\tau_0+1+\bar D(s_0)}\right)\,\hat s_2;
\end{eqnarray}
and
\begin{eqnarray}
\label{eigen2vp}
\nonumber
\hat s_2 = \sin (4\psi)\,\nu_{\infty}\,\hat g(bk)\,&&\left({\bar D(s_0)\,(s_y/s_0)^2\over \omega\,\tau_0+\bar D(s_0)}\right)\,\hat s_1\cr\\&& + \cos (4\psi)\,\nu_{\infty}\,\hat g(bk)\,\left({1\over \omega\,\tau_0+1+\bar D(s_0)}\right)\,\hat s_2.
\end{eqnarray}
From these we find, for $\psi=0$,
\begin{equation}
\omega(k,0)\,\tau_0 =\cases{\bar D(s_0)\,\Bigl[\nu_{\infty}\,\hat g(bk)\,(s_y/s_0)^2-1\Bigr]&($\hat s_1\,${\rm mode})\cr \\\left[1+\bar D(s_0)\right]\,\Bigl[\nu_{\infty}\,\hat g(bk)\,(s_y/s_0)^2-1\Bigr]&($\hat s_2\,${\rm mode});}
\end{equation}
and, for $\psi=\pi/4$,
\begin{equation}
\omega(k,\pi/4)\,\tau_0 =\cases{-\bar D(s_0)\,\Bigl[\nu_{\infty}\,\hat g(bk)\,(s_y/s_0)^2+1\Bigr]&($\hat s_1\,${\rm mode})\cr \\-\left[1+\bar D(s_0)\right]\,\Bigl[\nu_{\infty}\,\hat g(bk)\,(s_y/s_0)^2+1\Bigr]&($\hat s_2\,${\rm mode});}
\end{equation}
Now both modes are unstable at $\psi=0$, and both are stable at $\psi=\pi/4$. The instability occurs only in a finite range of stresses $1<(s_0/s_y)^2 <(\nu_{\infty}/\nu_c)$; thus, it can be suppressed by driving the system further into the viscoplastic regime by increasing the applied stress $s_0$.  Note that the geometry of this situation is similar to that of the crazing instability in polymers.  The pattern of most rapidly growing modes is aligned with the principal axes of the stress tensor rather than along the direction of maximum shear.

\section{Nonlinear Analysis}
\label{sec:Nonlinear}

We must look at nonlinear behavior in order to understand the implications of these linear instabilities.  To make progress along these lines without a great deal of new formal development, we can restrict ourselves to situations in which the principal axes of the stress tensor retain a single, fixed orientation everywhere throughout the system and at all times.  Within this restriction, we can examine the $\hat s_1$ mode that is unstable for $s_0<s_y$, because the applied and induced stresses in that case remain everywhere aligned along the $x,\,y$ axes, and all spatial variations occur only along the diagonal orientations $\psi=\pm \pi/4$.  While this mode is not the most strongly unstable, it is the one that couples to the applied stress and thus describes the deformations that would be measured in response to the applied loads. The restriction of fixed stress orientation substantially limits the conclusions that we can draw from this analysis. On the other hand, there is no longer any reason to use the quasilinear approximation.  Several physically important features of the following results  emerge from the strong stress dependence of the full STZ theory. 

It is convenient to make the following changes of variables
\begin{equation}
s_{xx}=-s_{yy}={1\over\gamma}\,\sigma(\xi,t);~~~~\Delta_{xx}=-\Delta_{yy}=\Delta(\xi,t),
\end{equation}
and $\xi=(x\pm y)/b\sqrt 2$; i.e. $\xi$ is the spatial variable  in units $b$ along one or the other of the diagonal directions. For simplicity, express time in units $R_0^{-1}$. With these transformations and the additional spatial dependence, Eqs.(\ref{deltadot}) and (\ref{lambdadot}) become
\begin{equation}
\label{deltadot2}
\dot\Delta = 2\,{\cal C}(\sigma)\,\left[{\cal T}(\sigma)\,\Lambda -\Delta\right]\,(1-\sigma\,\Delta)
\end{equation}
and
\begin{equation}
\label{lambdadot2}
\dot\Lambda = 2\,\sigma\,{\cal C}(\sigma)\,\left[{\cal T}(\sigma)\,\Lambda -\Delta\right]\, (1 - \Lambda).
\end{equation}
Here,
\begin{equation}
\label{Cdef2}
{\cal C}(\sigma)=\exp\left[-\alpha\,\cosh\left({\sigma\over\sigma_1}\right)\right]\, \cosh\left[\alpha\,\sinh\left({\sigma\over\sigma_1}\right)\right];
\end{equation}
\begin{equation}
\label{Tdef2}
{\cal T}(\sigma) =\tanh\left[\alpha\,\sinh\left({\sigma\over\sigma_1}\right)\right];
\end{equation} 
and $\alpha = V_0^*/v_f$, $\sigma_1 = \gamma\,\bar\mu$.  Symbols such as ${\cal C}(\sigma)$ now denote functionals ${\cal C}[\sigma(\xi,t)]$.  The yield stress $\sigma_y$, according to Eq.(\ref{yieldstress}), is determined by 
\begin{equation}
\label{yieldstress2}
\sigma_y\,{\cal T}(\sigma_y)= 1;
\end{equation}

Tightly packed, hard-to-deform materials such as glasses at low temperatures or the Lennard-Jones material studied in \cite{FL} have small free volumes $v_f$ and therefore large values of $\alpha$. Conversely, softer materials such as polymers or clays should have smaller values of $\alpha$. For large $\alpha$, ${\cal T}(\sigma)\cong 1$ at all values of $\sigma$ appreciably larger than $\sigma_1/\alpha$.  Thus, Eq.(\ref{yieldstress2}) implies that $\sigma_y$ is of order unity or, equivalently, $s_y$ is of order $\gamma^{-1}$. In \cite{FL}, the value of $\bar\mu$ was found to be of order the yield stress, roughly two orders of magnitude smaller than the elastic shear modulus. This estimate, along with the preceding estimate for $\gamma$, then implies that $\sigma_1\cong 1$. 

To be consistent with previous notation, write $\sigma(\xi,t) = \sigma_0(t)+\tilde\sigma(\xi,t)$, where $\sigma_0(t)$ is the applied stress and $\tilde\sigma$ is the induced stress fluctuation.  Because only the $\hat s_1$ mode plays a role in this situation, we can write Eq.(\ref{sdoteq}) in the form
\begin{equation}
\label{dotsigma}
\dot{\tilde\sigma}=\tilde\nu\,g*\dot\Delta,
\end{equation}
where $\tilde\nu = 2\,\mu\,\gamma\,(\Delta\varepsilon)\,b^2\,n_{\infty}$.  The quantity $2\,\mu\,\gamma$ is approximately the ratio of the shear modulus to the yield stress, and is therefore very roughly of order $10^2$.  The quantity $(\Delta\varepsilon)$ is of order unity.  As seen in the stability analysis, we need values of $\tilde\nu$ greater than about $0.25$ to generate instabilities. Accordingly,  $b^2\,n_{\infty}$ must be of order $10^{-2}$, which is consistent with our basic assumption that the STZ's are widely separated from each other. 

In Eq.(\ref{dotsigma}), the symbol $g*$ denotes the integral operator obtained by Fourier transforming $\hat g(\kappa)$:
\begin{equation}
g*\dot\Delta(\xi)= \,\int d\xi'\,\int {d\kappa\over2\pi}\,\hat g(\kappa)\,e^{i\kappa(\xi-\xi')}\,\dot\Delta(\xi').
\end{equation}
Using Eq.(\ref{ghat}), we find
\begin{equation}
g*\dot\Delta(\xi)=\int_{-1}^1 d\xi'\,g(\xi-\xi')\,\dot\Delta(\xi')
\end{equation}
where
\begin{equation}
\label{gxi}
g(\xi)=-{1\over 2\xi^3}\,\left[3\,\sin(2\theta_1)+3\,\sin(4\theta_1) + \sin(6\theta_1)\right];~~~~\theta_1=\arccos \xi.
\end{equation}
The function $g(\xi)$ is shown in Fig. 1.  It is strictly local; that is, it vanishes for $|\xi|>1$ (an artifact of the sharp cutoff in Eq.(\ref{sdoteq})).  The combination of Eqs.(\ref{deltadot2}), (\ref{lambdadot2}), and (\ref{dotsigma}) is a nonlinear system of equations that can be solved for $\Delta(\xi,t)$, $\Lambda(\xi,t)$, and $\sigma(\xi,t)$.  

Before starting this nonlinear analysis, however, note that the stability spectrum for the uniform state of the full STZ model is qualitatively different from the quasilinear result, Eq.(\ref{omega45}), in the case of hard materials with large $\alpha$.  For applied stress $\sigma_0 < \sigma_y$, and for modes of dimensionless wavenumber $\kappa$ oriented along one of the diagonals, the amplification rate is
\begin{equation}
\label{omegakappa}
\omega(\kappa)= 2\,{\cal C}(\sigma_0)\,\Biggl(\Bigl[1-\sigma_0\,\Lambda_0\,{\cal T}(\sigma_0)\Bigr]\,\Lambda_0\, {\cal T}'(\sigma_0)\,\tilde\nu\,\hat g(\kappa) - \Bigl[1-\sigma_0\,{\cal T}(\sigma_0)\Bigr]\Biggr).
\end{equation}
This formula has the same structure as the $\hat s_1$ mode in Eq.(\ref{omega45}).  However, the quantity $\tilde\nu\,\hat g(\kappa)$ is multiplied here by an extra factor ${\cal T}'(\sigma_0)$, which becomes exponentially small for $\sigma_0 > \sigma_1/\alpha$. Therefore, for large $\alpha$ and for $\sigma_0$ not too small, the system is stable except for unphysically large values of $\tilde\nu$.  The ``weakening'' factors in Eq.(\ref{omegakappa}) (in square brackets) also slow the instability at stresses near $\sigma_y$. At small values of $\sigma_0$, on the other hand, the factor $2\,{\cal C}(\sigma_0)$ becomes small of order $\exp(-\alpha)$, causing any instability that might occur there to grow very slowly.  We therefore expect to find little or no microstructural shear-banding instability in hard materials.  We shall see, however, that the absence of instability does not imply the absence of shear banding.   

Turn now to the numerical solutions of Eqs.(\ref{deltadot2}), (\ref{lambdadot2}), and (\ref{dotsigma}).  To solve these equations, I have used a simple discretization of the integral kernel $g(\xi)$, and have applied periodic boundary conditions, usually choosing the length of the system to be an integral number of wavelengths $2\pi/\kappa_c$ of the most rapidly growing mode. In this way, the problem is reduced to a large set of coupled, ordinary, nonlinear differential equations.  The initial conditions are always $\tilde\sigma(\xi,0)=0$ and $\Lambda(\xi,0)=\Lambda_{in}$. We shall see that $\Lambda_{in}$ provides crucial information about the way in which the system was prepared, specifically, the degree to which it was annealed. In order to allow an instability to develop, I have chosen $\Delta(\xi,0)$ to be some small-amplitude, random white noise, corresponding to uncorrelated inhomogeneities in the initial orientations of the STZ's.  Specifically,
\begin{equation}
<\Delta(\xi,0)\,\Delta(\xi',0)>=M\,\delta(\xi-\xi'),
\end{equation}
where the angular brackets denote a statistical average and $M$ is the noise strength.  

For reasons mentioned previously, which I shall discuss further in Section \ref{sec:Conclusion}, it is not sensible to try to make quantitative comparisons between the results of this highly simplified calculation and real experimental data.  Nevertheless, it is useful to use the results of Hasan and Boyce (HB)\cite{HB} as a qualitative guide.  First consider their creep tests in which a uniform applied stress $\sigma_0$ is turned on suddenly at $t = 0$ and held constant.  I assume that HB were using a relatively soft material, and therefore choose $\alpha= \sigma_1=1$. For these parameters, the yield stress is $\sigma_y= 1.132$. In their Fig. 4, HB show creep-test data for a sample that underwent a sharp transition from mean-field-STZ ``jammed'' behavior to rapid yielding over a narrow range of applied stresses, approximately 90 - 100 MPa.  From this behavior, I guess that their ultimate yield stress --- their analog of $\sigma_y$ --- was about 100 MPa, and that I should look at dimensionless stresses $\sigma_0$ in some range near unity in order to make comparisons with their results.  

The sample used in this particular HB experiment was annealed so that the initial population of transformation zones must have been quite small.  It follows that we should look at small values of $\Lambda_{in}$; that is, as in HB, we expect that the density of active sites is small at first and then increases as the system undergoes plastic deformation.  Our stability analysis already gives us useful information about how this may happen.  In Fig. 2, I have used Eq.(\ref{omegakappa}) to compute the amplification rate for the most rapidly growing wavelength, $\omega(\kappa_c)$, as a function of the stress $\sigma_0$. For lack of a better estimate, I have used $\tilde\nu=0.4$ here and throughout the rest of this presentation, because that value produces moderate instabilities and, as mentioned previously, is consistent with rough estimates of its constituent parameters. To compute $\Lambda_0$, I have used Eqs.(\ref{deltadot2}) and (\ref{lambdadot2}) to find
\begin{equation}
\label{lambdain}
\Lambda_0=\lim_{t\to\infty}\,\Lambda(t)= {\Lambda_{in}\over 1-(1-\Lambda_{in})\, \sigma_0\,{\cal T}(\sigma_0)}.
\end{equation}
The two functions $\omega(\kappa_c,\sigma_0)$ shown in the figure correspond respectively to a highly annealed specimen with $\Lambda_{in}=0.1$, and to one that is very rapidly quenched so that $\Lambda_{in}=1.0$.  The quenched specimen is strongly unstable at small stresses and stabilizes only just below $\sigma_y$.  In contrast, the annealed specimen is unstable only in a narrow range of stresses from $\sigma_0\cong 0.97$ up to the yield stress, $\sigma_y=1.132$.  At the smaller stresses, the density of zones never grows large enough to trigger the instability.  

The corresponding results of the nonlinear calculation for an annealed system with $\Lambda_{in}=0.1$ are shown in Fig. 3. Here, the plastic strain $\varepsilon^{pl}(t)$, in units $\varepsilon_0= b^2\, (\Delta\varepsilon) \,n_{\infty}/2$, is shown as a function of time (in units $R_0^{-1}$) for a sequence of applied stresses corresponding roughly to the stress range shown in HB Fig. 4.  Specifically, the stresses shown in Fig. 3 are $\sigma_0=0.4,\, 0.8,\, 0.9,\, 0.95,\, 0.975, \,1.00, \, 1.05, \,1.1,\, {\rm and}\, 1.2$. Again, $\tilde\nu = 0.4$. The noise strength is $M=10^{-5}$.  As expected, for stresses $\sigma_0 \le 0.95$, the strain rises to its uniform, steady-state value at which the system remains ``jammed'' indefinitely. There is a rapid transition in the range $0.95 < \sigma_0 < \sigma_y$, where the strain rate first slows but then increases to a new steady-state value, indicating the onset of microstructural shear banding.  In this range of stresses, the system enters a stable flowing regime in which the dimensionless shear rate
\begin{equation}
\label{qdef}
q(\xi,t)=2\,{\cal C}(\sigma)\,\left[{\cal T}(\sigma)\,\Lambda-\Delta\right]
\end{equation}
is localized in bands, as shown in Fig. 4 for the case $\sigma_0=1.0$ and time $t=100$. As the stresses $\sigma_0$ rise above $\sigma_y$, the banding instability becomes transient and the system reverts at long times to its mean-field STZ behavior for the viscoplastic regime.

Now consider the second kind of experiment mentioned in the Introduction, i.e. the conventional situation in which a stress-strain curve is measured at constant strain rate.  Assume that the spatially averaged total strain rate, say $\varepsilon_0\,q_0$, is held fixed. Assume further that the total strain is a simple sum of elastic and plastic parts, and that the elastic strain $\varepsilon^{el}$ is given by Hooke's law, i.e. $\sigma = 2\,\mu\,\gamma\,\varepsilon^{el}$.  Because the spatial average of the fluctuating stress $\tilde\sigma$ vanishes, the condition of constant $q_0$ becomes an equation of motion for the applied stress:
\begin{equation}
\label{qeqn}
\dot\sigma_0={1\over 2}\,\tilde\nu\,\left(q_0-\bar q(t)\right),
\end{equation}
where $\bar q(t)$ denotes the spatial average of the plastic strain rate $q(\xi,t)$ defined in Eq.(\ref{qdef}). 

Figure 5 shows stress-strain curves computed using Eqs.(\ref{deltadot2}), (\ref{lambdadot2}), and (\ref{dotsigma}), supplemented by Eq.(\ref{qeqn}). The material parameters are the same as in Fig. 3, in particular, $\Lambda_{in}=0.1$.  The strain rates are: $q_0=0.008,\, 0.016,\, 0.032,\,  0.048,\, {\rm and}\, 0.064$.  The peak stresses seen here are roughly comparable to those shown for the analogous HB experiments, and are in the transition range seen in the constant-stress calculations with a slight overshoot because the applied stress continues to rise during the onset of the instability.  The subsequent stress drops, however, are substantially larger and sharper than in HB.  I have not so far found any way of correcting this discrepancy simply by adjusting parameters in the present theory, and therefore suspect that the problem lies with one or more of the major simplifying assumptions that are built into it. 

The behaviors shown in Figs. 3 and 5 represent only a very small part of the  space spanned by the parameters $\alpha$, $\tilde\nu$, $\Lambda_{in}$, etc.. For example, the stability spectrum shown in Fig. 2 implies that creep tests performed on rapidly quenched soft materials, with large $\Lambda_{in}$, would exhibit shear banding instabilities at arbitrarily small applied stresses, and that these instabilities would actually weaken and disappear at larger stresses.  Similarly, such materials (if they exist) would exhibit strain softening at constant strain rate only at low rates and with small peak stresses.  It will be interesting to learn whether such behaviors, or others that may be less apparent results of this nonlinear model, actually occur in nature.

The last question that I shall address with these techniques is whether there may be some connection between the microstructural shear banding discussed so far and the formation of macroscopic shear bands in materials that do not soften in this manner. Consider a case in which the system is strongly stable against formation of microstructural shear bands, but in which there is a weak spot characterized by a locally high density of STZ's. Further, let the orientations of the zones in this weak spot be such that they will transform easily in response to an applied stress, and thus will induce additional stresses in their neighborhoods.  

The behavior of such a system is shown in Fig. 6.  Here, as before, $\sigma_1=1$, and $\tilde\nu = 0.4$, but $\alpha=10$.  Initially, $\Delta(\xi,0)=0$ and $\Lambda(\xi,0)=0.5$ everywhere except in a narrow region (9 of 420 discretization intervals) at $\xi = 3.5$ where $\Delta(\xi,0)=-1,~~\Lambda(\xi,0)=2$.  There is no other heterogeneity and no noise in this initial state.  A constant, uniform stress $\sigma_0=0.5$ is applied abruptly at $t=0$.  In the absence of the weak spot, this system would relax stably to a uniform, jammed state with $q=0$ after about 2000 time units.  (The relaxation is slower than in previous examples because of the large value of $\alpha$ and therefore small value of ${\cal C}(\sigma_0)$.)  Figure 6 shows the resulting dimensionless strain rate $q(\xi,t)$ at times $t= 200$ and $t=5000$.  At the earlier time, a central shear band and two weaker side bands are emerging, and the whole system is also deforming as indicated by the nonzero values of $q$ well away from the bands.  At the later time, all of the shear is taking place in the bands, and the peak strain rate in the central band is now an order of magnitude larger than it was earlier.  This configuration seems to be completely stable; I have continued the calculation out to $t=10,000$ and find no further changes. 

This numerical experiment is as close as I can come to simulating a shear band in this one-dimensional truncation of the plasticity problem.  In a real situation, the band would start at a stress concentration near some defect or surface irregularity, and would propagate away from that point along a direction of large shear stress.  (See, for example, observations by Kramer \cite{KRAMER}.) Such intrinsically two or three dimensional behavior has no analog here.  The most that can be said so far is that this calculation indicates the possibility of isolated band formation even in hard materials where no extended, microstructural shear-band patterns can occur.  It also suggests that a macroscopic shear band might consist of a cluster of microstructural bands. 

\section{Concluding Remarks}
\label{sec:Conclusion}

The STZ theory, including its extension discussed here, exhibits a wide range of behaviors which look qualitatively like phenomena that occur in real materials.  In contrast to other more phenomenological theories of plasticity, it has the advantage of being based on a microscopic picture of the internal states of noncrystalline solids; thus, it ultimately should be possible to compute the STZ constitutive parameters from first principles.  More important, in my opinion, is the fact that the STZ theory consists of a small set of equations of motion for internal state variables, and therefore is intrinsically simpler and more general than conventional theories\cite{LUBLINER,HILL}.  The structure of the STZ theory, by itself and without reference to specific mechanisms, has experimental implications.  For example, the diverging relaxation time near the ultimate yield stress in non-softening materials is a very general prediction that ought to be checked experimentally.  Another  possibility, as I have argued in a recent publication, is that the new dynamical degrees of freedom in the STZ model may resolve the puzzle of how breaking stresses can be transmitted through plastic zones at the tips of brittle cracks.\cite{Lfract}

Microstructural shear banding, while not so general a prediction as the diverging relaxation time near the yield stress, seems to be an inevitable consequence of the specific STZ mechanism; the banding instability comes directly from the quadrupolar symmetry of the STZ transformation.  However, the theoretical analysis presented here is still far from definitive.  Its most serious shortcoming is that it is only one dimensional in the places where it tries to relate to experimental observations, i.e. in Section \ref{sec:Nonlinear}.  A more likely picture is the one shown in Fig. 6 of Bulatov and Argon \cite{BA1}, where the two-dimensional pattern of shear bands consists of finite-sized patches oriented along the two equivalent directions of maximum shear stress.  The way in which these patches interact and interfere with each other may be a crucial aspect of the pattern-forming mechanism, in which caase a fully two-dimensional analysis will be necessary in order to understand the shear-banding instability in the STZ theory.  

There are several other serious shortcomings of the present analysis, most --- or all --- of which I believe can be remedied by plausible extensions of the existing theory.  One such shortcoming is that the STZ theory assumes that there is only a single kind of zone rather than a distribution of them with a range of transformation rates.  Hasan and Boyce\cite{HB} include such a distribution in their constitutive model, and they use its width as one of their adjustable parameters for fitting experimental data.  I suspect that the implicit assumption of an infinitely sharp distribution in the present analysis may be the principal reason why the stress drops in Fig. 5 are so abrupt and deep.  It would not be difficult to rewrite the present theory as an average over different varieties of STZ's and thus smooth out such unrealistically sharp features. However, it would be much more satisfactory to derive such distributions from a first-principles theory of deformations in amorphous materials rather than just to introduce them phenomenologically.

Another shortcoming is that I have not, so far, extended the STZ theory to finite temperatures.  Including thermal activation in the rate factors seems a simple matter, and the temperature dependence of the phenomena studied here might provide useful tests of the basic theoretical ideas.

Finally, there is the question of length scales in this theory. It seems puzzling that the only length scale to emerge in this analysis is $b$, the size of the zones.  Why should the spacing of the microstructural shear bands scale simply with $b$ rather than depending also on the density $n_{tot}$?  Why, for that matter, should such an instability occur only on microscopic rather than macroscopic length scales?  Perhaps a better theory will produce a screening length for the interactions between zones.  And perhaps, in a two-dimensional version of the theory, interactions between patches of different orientations will produce a characteristic patch size.  There seems to be a wide range of possibilities both for extending the STZ picture and for testing its general validity.     

\acknowledgments

I would like to thank Sharad Ramanathan for first suggesting how the STZ mechanism might produce a shear-banding instability, and Ali Argon and Ed Kramer for useful discussions about the mechanical properties of amorphous materials.  I also thank Daniel Lavallee and Anthony Foglia for important help with the numerical analysis, and Leonid Pechenik and Daniel Rabinowitz for pointing out errors in earlier versions of the manuscript.  

This research was supported primarily by U.S. Department of Energy Grant DE-FG03-99ER45762.  It was also supported in part by the MRSEC Program of the NSF under award number DMR96-32716 and by a grant from the Keck Foundation for Interdisciplinary Research in Seismology and Materials Science.

\clearpage

\begin{figure}[t]
\epsfxsize=3.5in
\centerline{\epsffile{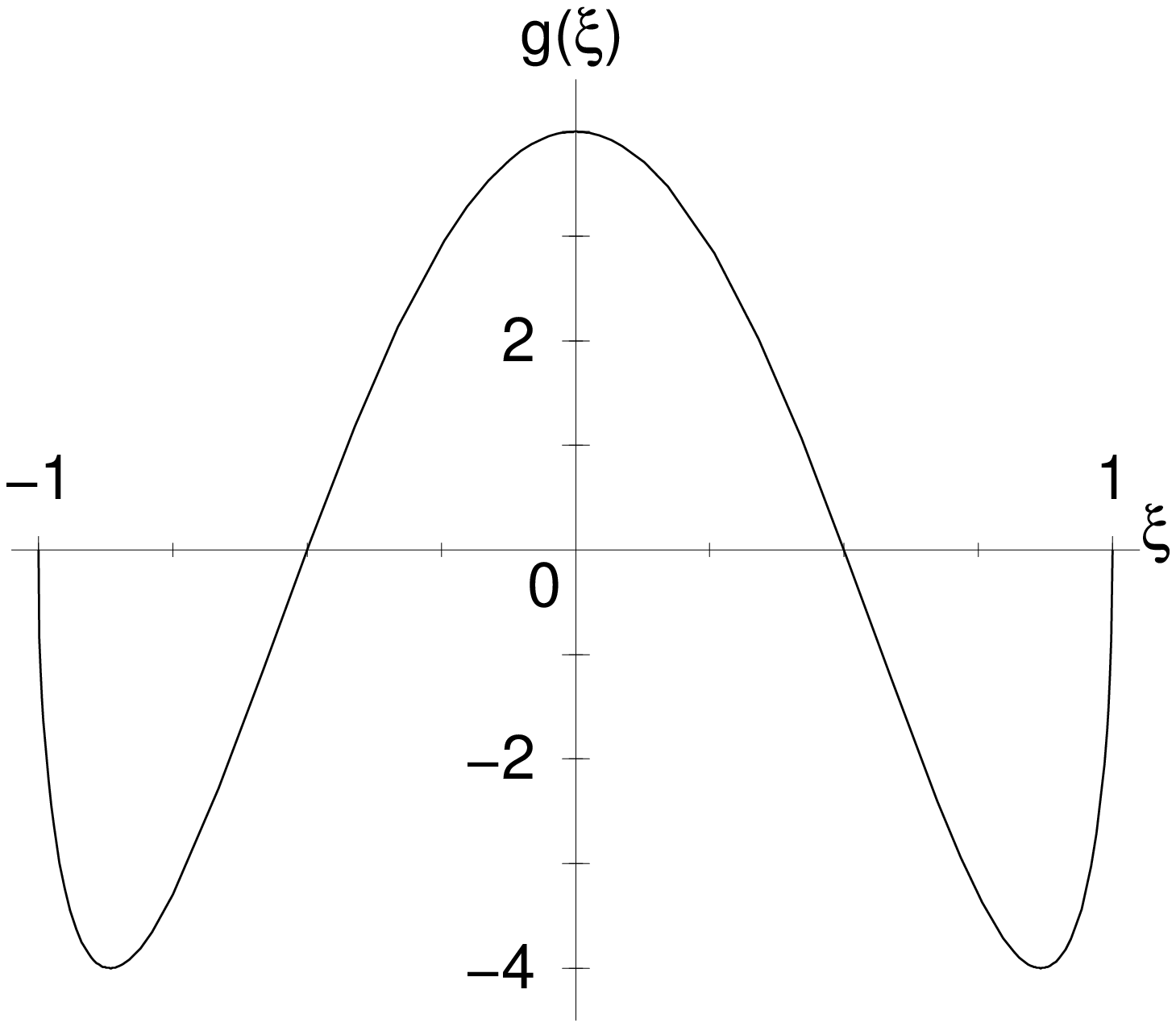}}
\caption{Integral kernel $g(\xi)$ defined in Eq.(\ref{gxi}).}
\end{figure}

\begin{figure}[t]
\epsfxsize=4in
\centerline{\epsffile{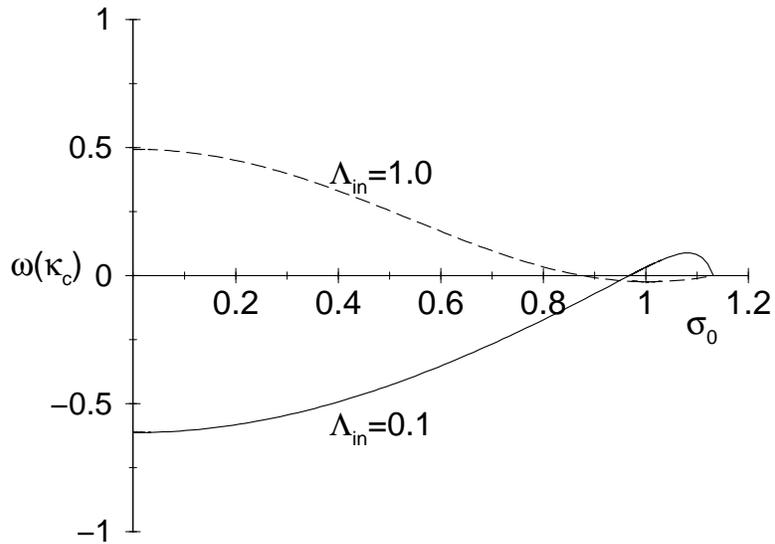}}
\caption{Amplification rate $\omega(\kappa_c)$ defined in Eq.(\ref{omegakappa}), as a function of applied stress $\sigma_0$, for two different values of $\Lambda_{in}$ according to Eq.(\ref{lambdain}). The solid curve, for $\Lambda_{in}=0.1$, corresponds to an annealed system with few defects; the dashed curve, for $\Lambda_{in}=1.0$, corresponds to a system that has been quenched so rapidly that the initial STZ density is high.}
\end{figure}

\begin{figure}[t]
\epsfxsize=4in
\centerline{\epsffile{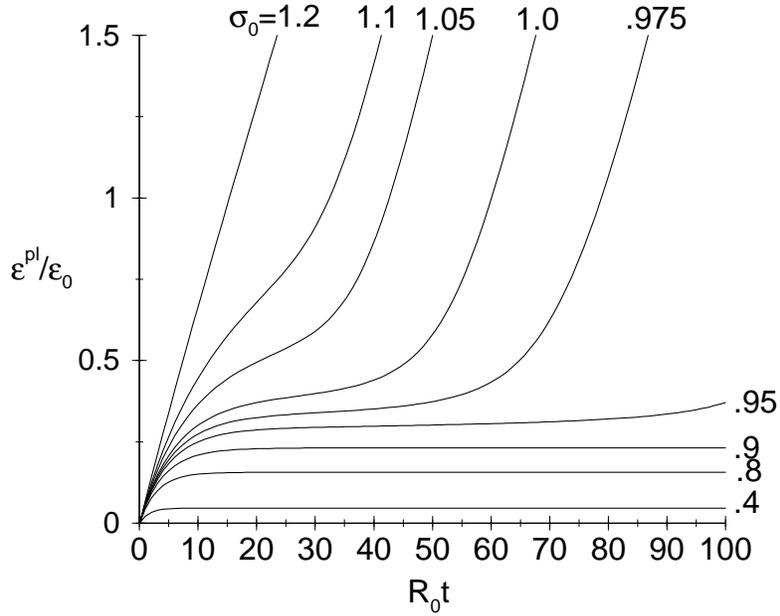}}
\caption{Plastic strain as a function of time for a sequence of simulated constant-stress experiments (creep tests). The constitutive parameters are $\alpha=1$, $\sigma_1=1$, and $\tilde\nu = 0.4$. The initial (dimensionless) density of STZ's is $\Lambda_{in}=0.1$; and the sequence of applied stresses $\sigma_0$ is as shown.  The mesh size for numerical discretization is $\Delta\xi = 0.05$. Stresses $\sigma_0$ are normalized so that the ultimate yield stress is $\sigma_y= 1.132$. Strains are in units $\varepsilon_0$, as defined in Eq.(\ref{parameterdefs}). Times $t$ are in units $R_0^{-1}$, as defined in Eq.(\ref{ratefactor}).}
\end{figure}

\clearpage

\begin{figure}[t]
\epsfxsize=4in
\centerline{\epsffile{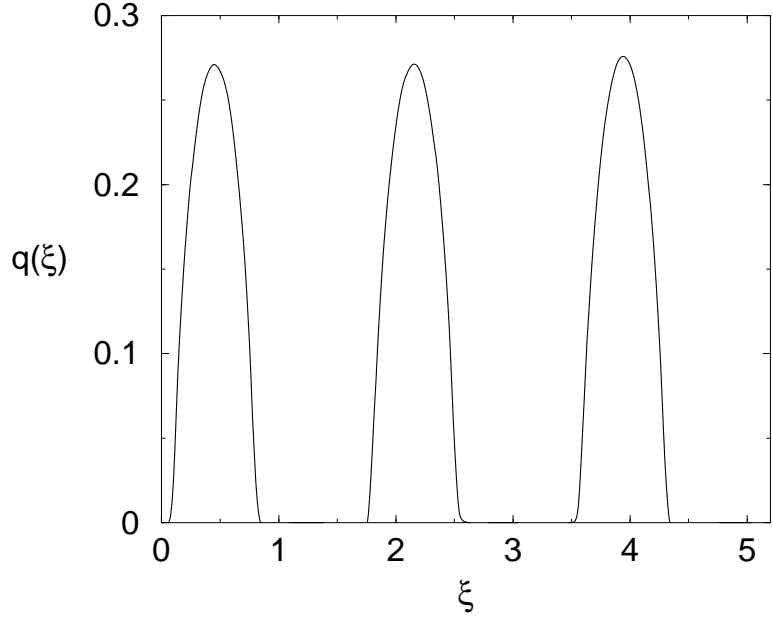}}
\caption{Strain rate $q$, as a function of position $\xi$, corresponding to the point $t=100$ on the curve for $\sigma_0=1.0$ in Fig. 3. $q$ is in units $\varepsilon_0\,R_0$.}
\end{figure}

\begin{figure}[t]
\epsfxsize=4in
\centerline{\epsffile{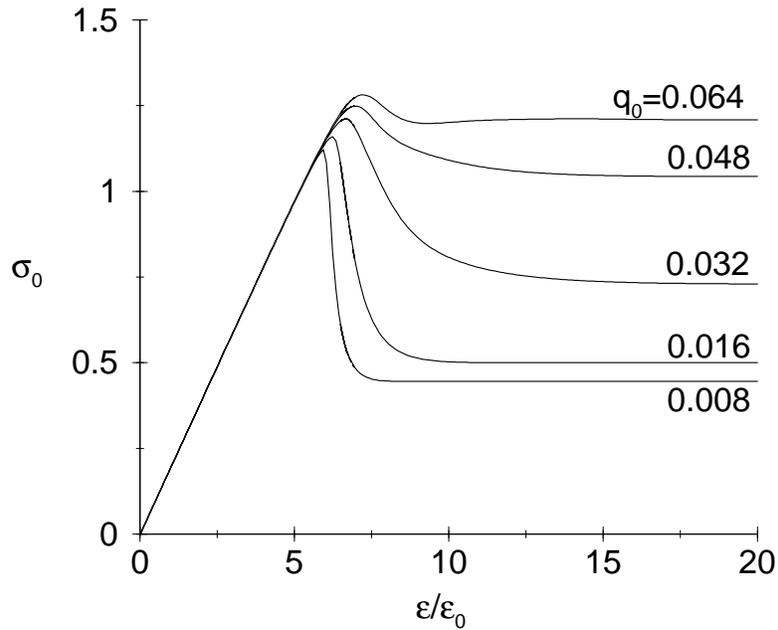}}
\caption{Stress-strain curves for a sequence of simulated constant strain-rate experiments. All material parameters are the same as in Fig. 3.  The strain rates $q_0$ are as shown.}
\end{figure}

\begin{figure}[t]
\epsfxsize=3.5in
\centerline{\epsffile{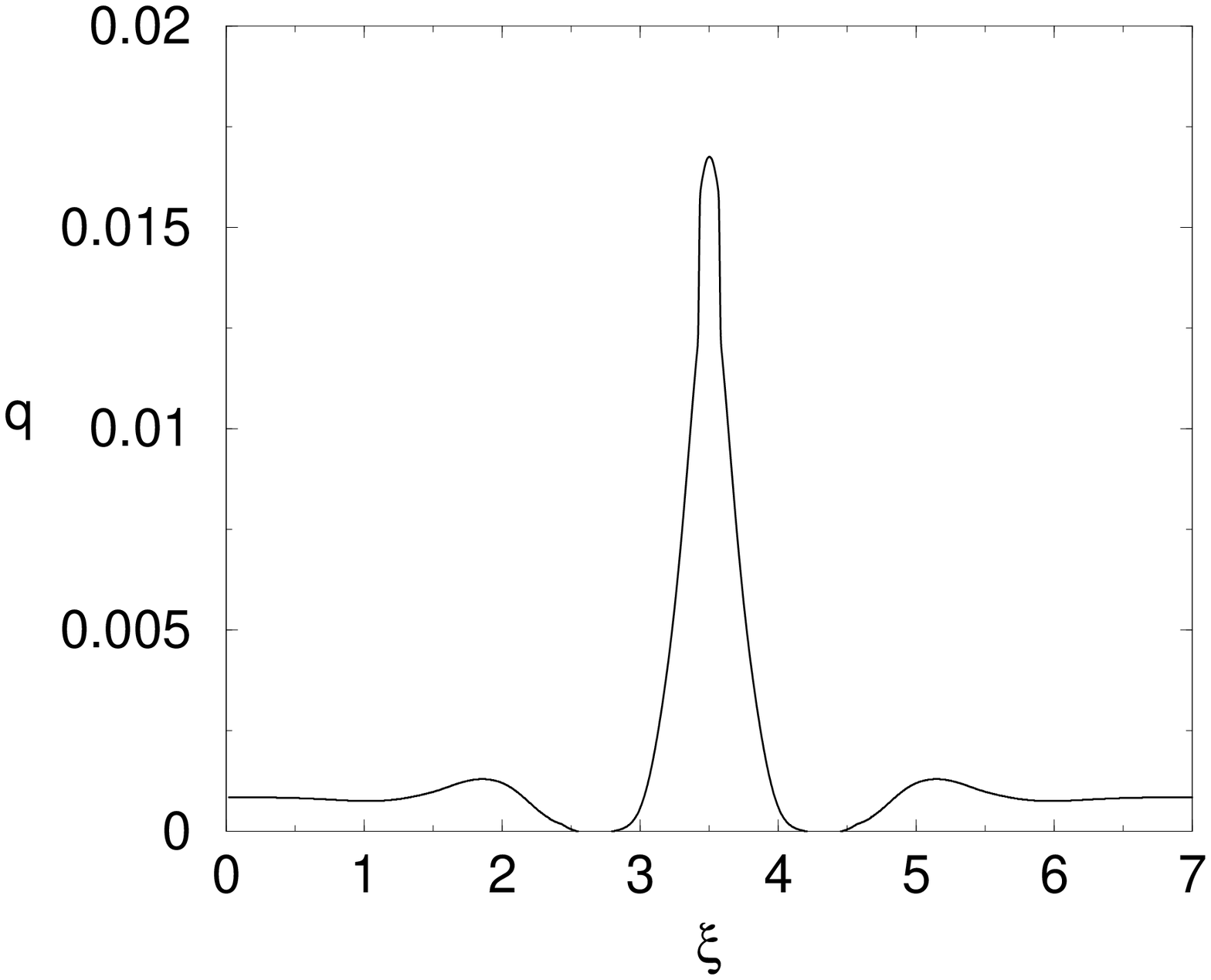}}
\end{figure}

\begin{figure}[t]
\epsfxsize=3.5in
\centerline{\epsffile{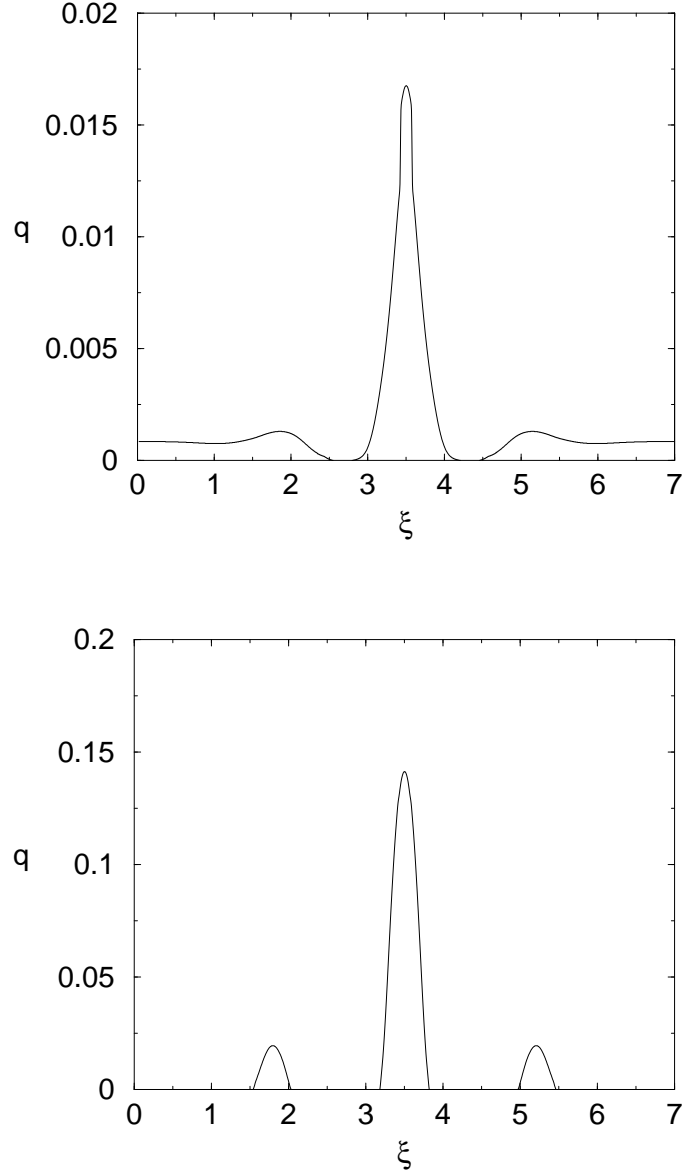}}
\caption{Two graphs of the strain rate $q$ as a function of position $\xi$ showing the emergence of shear bands near a weak spot in an otherwise stable system. The constitutive parameters are $\alpha=10$, $\sigma_1=1$, and $\tilde\nu = 0.4$. The uniform applied stress is $\sigma_0 = 0.5$. Initially, $\Lambda_{in}=0.5$, $\Delta = 0$ everywhere except in a very narrow region near $\xi = 3.5$ where,  at $t=0$, $\Lambda_{in}=2.0$, $\Delta=-1$.  In (a), where $t=200$, the system as a whole is still deforming in response to the applied stress, and the central shear bands are just beginning to appear.  In (b), where $t=5000$, all of the deformation rate is concentrated in the bands, and the whole system has reached its steady state configuration.}
\end{figure}

\end{document}